\documentclass[aps,prd,twocolumn,groupedaddress,showpacs,showkeys,superscriptaddress,nofootinbib]{revtex4}%
\usepackage{graphics}%
\usepackage{amsfonts}%
\usepackage{mathrsfs}     

\begin{document}%
\def\Barcelo{Barcel\'o}%
\title{Fate of gravitational collapse in semiclassical gravity}%
\author{Carlos \Barcelo}%
\affiliation{Instituto de Astrof\'{\i}sica de Andaluc\'{\i}a, CSIC,
Camino Bajo de Hu\'etor 50, 18008 Granada, Spain}%
\author{Stefano Liberati}%
\affiliation{International School for Advanced Studies, Via Beirut
2-4, 34014 Trieste, Italy and INFN, Sezione di Trieste}%
\author{Sebastiano Sonego}%
\affiliation{Universit\`a di Udine, Via delle Scienze 208, 33100 Udine, Italy}%
\author{Matt Visser}%
\affiliation{School of Mathematics, Statistics, and Computer Science,
Victoria University of Wellington, New Zealand}%
\date{\today}%
\bigskip%
\begin{abstract}%

While the outcome of gravitational collapse in \emph{classical\/}
general relativity is unquestionably a black hole, up to now no full
and complete \emph{semiclassical\/} description of black hole
formation has been thoroughly investigated. Here we revisit the
standard scenario for this process. By analyzing how semiclassical
collapse proceeds we show that the very formation of a trapping
horizon can be seriously questioned for a large set of, possibly
realistic, scenarios. We emphasise that in principle the theoretical
framework of semiclassical gravity certainly allows the formation of
trapping horizons. What we are questioning here is the more subtle
point of whether or not the standard black hole picture is
appropriate for describing the end point of {\em realistic\/}
collapse. Indeed if semiclassical physics were in some cases to
prevent  formation of the trapping horizon, then this suggests the
possibility of new collapsed objects which can be much less
problematic, making it unnecessary to confront the information
paradox or the run-away end point problem.

\end{abstract}%
\maketitle%

\def\e{{\mathrm e}}%
\def\g{{\mbox{\sl g}}}%
\def\Box{\nabla^2}%
\def\d{{\mathrm d}}%
\def\R{{\rm I\!R}}%
\def\ie{{\em i.e.\/}}%
\def\eg{{\em e.g.\/}}%
\def\etc{{\em etc.\/}}%
\def\etal{{\em et al.\/}}%
\newcommand{\scri}{\mathscr{I}}
\def\HRULE{{\bigskip\hrule\bigskip}}%
\def\implies{{\Rightarrow}}

\section{Introduction}

Although the existence of astrophysical black holes is now commonly
accepted, we still lack a detailed understanding of several aspects
of these objects. In particular, when dealing with quantum field
theory in a spacetime where a classical event horizon forms, one
encounters significant conceptual problems, such as the
information-loss paradox linked to black hole thermal
evaporation~\cite{Hawking:science, hawking-ter, hawking-paradox,
preskill}.

The growing evidence that black hole evaporation may be compatible
with unitary evolution in string-inspired scenarios (see, \eg,
reference~\cite{AdS})\footnote{See, however, a recent article by
D.~Amati~\cite{Amati:2006fr} for an alternative point of view on the
significance of these results.} has in recent years led to a revival
of interest in, and extensive modification of,
early~\cite{Roman-Bergmann}  alternative semiclassical scenarios for
the late stages of gravitational collapse~\cite{hayward,
ashtekar-bojowald}. (See also~\cite{tipler,bardeen,york}.) Indeed,
while it is by now certain that the outcome of a realistic  {\em
classical\/} collapse is necessarily a standard black hole delimited
by an event horizon (that is, a region ${\cal B}$ of the total
spacetime $\cal M$ which does not overlap with the causal past of
future null infinity: ${\cal B}={\cal M}-J^{-}(\scri^+)\neq
\emptyset$), it has recently been suggested that only apparent or
trapping horizons might actually be allowed in nature, and that
somehow \emph{semiclassical\/} or \emph{quantum
gravitational\/}~\cite{ashtekar-bojowald, hawking-info, mathur}
effects could prevent the formation of a (strict, absolute) event
horizon,\footnote{``The way the information gets out seems to be
that a true event horizon never forms, just an apparent horizon''.
(Stephen Hawking in the abstract to his GR17
talk~\cite{hawking-info}.)} and hence possibly evade the necessity
of a singular structure in their interior.

Note that Hawking radiation would still be present, even in the
absence of an event horizon~\cite{hajicek, essential}. Moreover, the
present authors have noticed that, kinematically, a collapsing body
could still emit a Hawking-like Planckian flux even if no horizon
(of any kind) is ever formed at any finite
time~\cite{quasi-particle-prl};\footnote{Recently, it was brought to
our  attention that this possibility was also pointed out in a paper
by P. Grove~\cite{grove}.} all that is needed being an exponential
approach to apparent/trapping horizon formation in infinite time.
Since in this case the evaporation would occur in a spacetime where
information by construction  \emph{cannot\/} be lost or trapped,
there would be no obstruction in principle to its recovery by
suitable measurements of quantum correlations. (The evaporation
would be characterized by a Planckian spectrum and not by a truly
thermal one.)

Inspired by these investigations we wish here to revisit the basic
ideas that led in the past to the standard scenario for
semiclassical black hole formation and evaporation. We shall see
that, while the formation of the trapping horizon (or indeed most
types of horizon) is definitely permitted in semiclassical gravity,
nonetheless the actual occurrence or non-occurrence of a horizon
will  depend delicately on the specific dynamical features of the
collapse.

Indeed, we shall argue that in realistic situations one may have
alternative end points of semiclassical collapse which are quite
different from black holes, and intrinsically semiclassical in
nature. Hence, it may well be that the compact objects that
astrophysicists currently identify as black holes correspond to a
rather different physics. We shall here suggest such an alternative
description by proposing a new class of compact objects (that might
be called ``black stars'') in which no horizons (or ergoregions) are
present.\footnote{These ``black stars'' are nevertheless distinct
from the recently introduced ``gravastars''~\cite{gravastar}.} The
absence of these features would make such objects free from some of
the daunting problems that plague black hole physics.

\section{Semiclassical collapse: The standard scenario}%

Let us begin by revisiting the standard semiclassical scenario for
black hole formation.  For simplicity, in this paper we shall
consider only non-rotating, neutral, Schwarzschild black holes;
however, all the discussion can be readily generalized to other
black hole solutions.

Consider a star of mass $M$ in hydrostatic equilibrium in empty
space. For such a configuration the appropriate quantum state is
well known to be the Boulware vacuum state $|0_{\rm
B}\rangle$~\cite{boulware}, which is defined unambiguously as the
state with zero particle content for static observers, and is
regular everywhere both inside and outside the star (this state is
also known as the static, or Schwarzschild,
vacuum~\cite{birrell-davies}). If the star is sufficiently dilute
(so that the radius is very large compared to $2M$), then the
spacetime is nearly Minkowskian and such a state will be virtually
indistinguishable from the Minkowski vacuum. Hence, the expectation
value of the renormalized stress-energy-momentum tensor (RSET) will
be negligible throughout the entire spacetime. This is the reason
why, when calculating the spacetime geometry associated with a
dilute star, one only needs to care about the classical contribution
to the stress-energy-momentum tensor (SET).

Imagine now that, at some moment, the star begins to collapse. The
evolution proceeds as in classical general relativity, but with some
extra contributions as spacetime dynamics will also affect the
behaviour of any quantum fields that are present, giving place to
\emph{both} particle production and additional vacuum polarization
effects.  Contingent upon the standard scenario being correct, if we
work in the Heisenberg picture there is a single globally defined
regular quantum state $|C\rangle =|\mathrm{collapse}\rangle$ that
describes these phenomena.

For simplicity, consider a massless quantum scalar field and
restrict the analysis to spherically symmetric solutions.  Every
mode of the field can (neglecting back-scattering) be described as a
wave coming in from $\scri^-$ (\ie, from $r\to +\infty$, $t\to
-\infty$), going inwards through the star till bouncing at its
center ($r=0$), and then moving outwards to finally reach $\scri^+$.
As in this paper we are going to work in $1+1$ dimensions (\ie, we
shall ignore any angular dependence), for later notational
convenience instead of considering wave reflections at $r=0$ we will
take two mirror-symmetric copies of the spacetime of the collapsing
star glued together at $r=0$ (see Fig.~\ref{F:diamond}). In one copy
$r$ will run from $-\infty$ to $0$, and in the other from $0$ to
$+\infty$. Then one can concentrate on how the modes change on their
way from $\scri^-_\mathrm{left}$ (\ie, $r\to -\infty$, $t\to
-\infty$) to $\scri^+_\mathrm{right}$ (\ie, $r\to +\infty$, $t\to
+\infty$). Hereafter, we will always implicitly assume this
construction and will not explictly specify ``left'' and ``right''
except  where it might cause confusion.
%
\begin{widetext}
\begin{center}
\begin{figure}[!htb]%

\vbox{ \hfil \scalebox{0.50}[0.50]{{\includegraphics{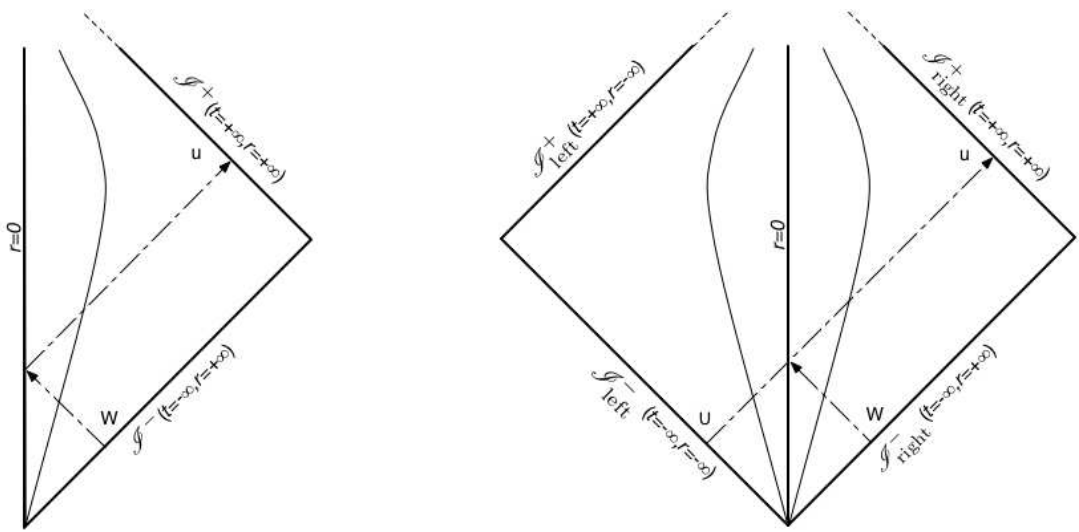}}}\hfil}%

\bigskip%
\caption{Standard conformal diagram for a collapsing star, and its
mirror-symmetric version.}
\label{F:diamond}%

\end{figure}%
\end{center}
\end{widetext}
%

Now, one can always write the field operator as
\begin{eqnarray}%
\widehat \phi(t,r)=\int {\mathrm d}\Omega \left[ \widehat
a_{\Omega} \; \varphi_{\Omega}(t,r) + \widehat
a_{\Omega}^\dagger \; \varphi_{\Omega}^*(t,r)
\right]\;,%
\end{eqnarray}%
where $\varphi_{\Omega}$ are the modes that near $\scri^-$
behave asymptotically as\footnote{We work in natural units.}
\begin{eqnarray}%
\varphi_{\Omega}(r,t) \approx {1 \over
(2\pi)^{3/2}(2\Omega)^{1/2} \,|r|} \; \e^{-i\Omega U}\;,%
\label{phi}%
\end{eqnarray}%
with $U=t-r$ and $\Omega>0$. One can then identify the state
$|C\rangle$ as the one that is annihilated by the destruction
operators associated with these modes: $\widehat a_{\Omega} |C
\rangle=0$. (One could also expand $\widehat\phi$ using a wave
packet basis~\cite{hawking-ter}, which is a better choice if one
wants to deal with behaviour localized in space and time.) Since the
spacetime outside the star is isometric with a corresponding portion
of Kruskal spacetime, and is static in the far past, the modes
$\varphi_\Omega$ have the same asymptotic expression as the Boulware
modes~\cite{boulware} near $\scri^-$ (\ie, for $t\to -\infty$).
Hence $|C\rangle$, the quantum state corresponding to the physical
collapse,  is (near $\scri^-$) indistinguishable from the Boulware
vacuum $|0_{\rm B}\rangle$. (But this will of course no longer be
true as one moves significantly away from $\scri^-$.)

Now, the semiclassical collapse problem consists of studying the
evolution of the geometry as determined by the semiclassical
Einstein equations
\begin{eqnarray}%
G_{\mu\nu}=8\pi\left(T^c_{\mu\nu}+\langle C|\widehat T_{\mu\nu}| C \rangle\right)\;,
\label{sceq}%
\end{eqnarray}%
where $T^c_{\mu\nu}$ is the classical part of the SET. Significant
deviations from the classical collapse scenario can appear only if
the RSET in equation (\ref{sceq}) becomes comparable with the
classical SET. In this analysis there are (at least) two important
results from the extant literature that have to be taken into
account:
\begin{itemize}%
\item
If a quantum state is such that the singularity structure of the
two-point function is initially of the Hadamard form, then Cauchy
evolution will preserve this feature~\cite{FSW}, at least up to the
edge of the spacetime (which might be, for instance, a Cauchy
horizon~\cite{HawBook}). The state $|C\rangle$ certainly satisfies
this Hadamard condition at early times~\cite{fnw}, hence it must
satisfy it also in the future, even if a trapping/event horizon
forms. (A trapping/event horizon is not a Cauchy horizon, and is not
an obstruction to maintaining the Hadamard condition.) As a
consequence of this fact the RSET cannot become singular anywhere on
the collapse geometry, independently of whether or not a
trapping/event  horizon is formed.\footnote{It is important to
understand exactly what this theorem does and does not say: If we
work in a well-behaved coordinate system (where the matrix of metric
coefficients is nonsingular and has finite components), then the
coordinate components of the RSET are likewise finite. But note that
finite does not necessarily imply small.}
\item For specific semiclassical models of the collapsing star it
has been numerically demonstrated (modulo several important
technical caveats) that the value of the RSET remains negligibly
small throughout the entire collapse process, including the moment
of horizon formation~\cite{pp}.\footnote{Similar results were, after
some discussion, found in $(1+1)$-dimensional models based on
dilaton gravity~\cite{2d}.} Subsequently, in this scenario quantum
effects manifest themselves via the slow evaporation of the black
hole.
\end{itemize}%

Thus in this standard scenario nothing prevents the formation of
trapped regions (or trapped/apparent/event horizons).  Given that
quantum-induced violations of the energy conditions~\cite{visser,
cosmo99} are taken to be small enough at this stage of the collapse,
one can still use Penrose's singularity theorem to argue that a
singularity will then tend to form. Assuming that quantum gravity
effects will not conspire to avoid this conclusion, then, in
conformity with all extant calculations and the cosmic censorship
conjecture, a spacelike singularity and a true event horizon will
form. The collapsed star settles down in a quasi-static black hole
and then ultimately evaporates.

This last feature can be easily derived
by considering an expansion of the field in a basis which
contains modes that near $\scri^+$ (\ie, for $r\to +\infty$, $t\to
+\infty$), behave asymptotically as%
\begin{eqnarray}%
\psi_\omega(r,t) \approx {1 \over
(2\pi)^{3/2}(2\omega)^{1/2} \, r} \; \e^{-i\omega u}\;,%
\label{psi}%
\end{eqnarray}%
with $u=t-r$ and $\omega>0$, so defining creation and annihilation
operators that differ from those associated with the modes
$\varphi_\Omega$ of equation (\ref{phi}). In a static configuration
a (spherical) wave coming from $\scri^-$ is blue-shifted on its way
towards the center of the star, and is then equally red-shifted on
its way out to $\scri^+$, arriving there undistorted.  However, in a
dynamically collapsing configuration the red-shift exceeds the
blue-shift, so that an initial wave at $\scri^-$ is distorted once
it reaches $\scri^+$.  In this sense the dynamical spacetime acts as
a ``processing machine'' for the normal modes of the field.
Expanding the distorted wave in terms of the undistorted basis at
$\scri^+$ tells us the amount of particle creation due to the
dynamics.  In particular one can take a wave packet centered on
frequency $\Omega$ on $\scri^-$ and ask what its typical frequency,
say $\omega$, will be when it arrives on $\scri^+$. The Bogoliubov
coefficients that allow us to express the annihilation operators
related to the modes (\ref{psi}) in terms of the creation operators
pertaining to the modes (\ref{phi}) are related to the number of
particles seen by asymptotic observers on $\scri^+$, which is
nothing else than the thermal flux of Hawking
radiation~\cite{Hawking:science, hawking-ter, birrell-davies}.

This can be rephrased saying that the physical state $|C\rangle$
corresponding to the collapse behaves like the Unruh vacuum $|0_{\rm
U}\rangle$~\cite{unruh} of Kruskal spacetime near the event horizon,
$\mathcal{H}^+$, and near $\scri^+$ (\ie, for $t\to +\infty$).
Indeed, in the Kruskal spacetime the Unruh state $|0_{\rm U}\rangle$
is a zero-particle state for a freely-falling observer crossing the
horizon, and corresponds to a thermal flux of particles at the
Hawking temperature for a static observer at
infinity~\cite{birrell-davies, scd}.  Given that at late times
classical black holes generated via classical gravitational collapse
are virtually indistinguishable from eternal black holes (see, for
instance,  the classical theorem in~\cite{wald}), the Unruh vacuum
is the only quantum state on Kruskal spacetime which appropriately
(near $\scri^+$ and $\mathcal{H}^+$) simulates the physical vacuum
in a spacetime with an event horizon formed via gravitational
collapse.

However as previously mentioned, this standard scenario leads to
several well known problems (or at the very least, disquieting
features):
\begin{itemize}%
\item Modes corresponding to quanta detected at $\scri^+$ have an
arbitrarily high frequency on $\scri^-$ (this is the so-called
trans--Planckian problem~\cite{unruh}).
\item The run-away end point of the evaporation process (the
Hawking temperature is inversely proportional to the black hole
mass) prevents any well-defined semiclassical answer regarding the
ultimate fate of a black hole~\cite{Hawking:science}.
\item If eventually the black hole completely evaporates, leaving
just thermal radiation in flat spacetime, then it would seem that
nothing would prevent a unitarity-violating evolution of pure states
into mixed states, contradicting a basic tenet of (usual) quantum
theory (this is one aspect of the so-called information-loss
paradox~\cite{hawking-paradox, preskill}). Such a difficulty for
reconciliating quantum mechanics with general relativity seems to
persist even when imagining many alternative scenarios for the end
point of the evaporation, so that one can still continue to talk
about an information-loss problem~\cite{hawking-paradox, preskill}.
\end{itemize}%
All in all, it is clear that this semiclassical collapse scenario is
evidently plagued by significant difficulties and obscurities that
still need to be understood. For this reason we think it is
worthwhile to step back to a clean slate, and to revisit the above
story uncovering all the hidden assumptions.

\section{Semiclassical collapse: A critique}%

It is easy to argue that one cannot trust a semiclassical gravity
analysis once a collapsing configuration has entered into a
high-curvature (Planck-scale) regime; this is expected in the
immediate neighborhood of the region in which the classical
equations predict the appearence of a curvature singularity.  Once
the formation of a trapped region is \emph{assumed\/}, any solution
of the problems  mentioned above seems (naively) to demand an
analysis in a full-fledged theory of quantum gravity. Here, however,
we are questioning the very formation of a trapped region in
astrophysical collapse. In analyzing this question we will see that
semiclassical gravity provides a useful and sensible starting point.
Moreover, we will also show that it provides some indications as to
how the standard scenario might be modified.

\subsection{The trans--Planckian problem}

One potential problem with the semiclassical gravity framework, when
used to analyze the onset of horizon formation, is the
trans--Planckian problem. While this problem is usually formulated
in static spacetimes, for our purposes we wish to look back to its
origin in a collapse scenario.

We can, as usual, encode the dynamics of the geometry in the
relation $U=p(u)$ between the affine null coordinates $U$ and $u$,
regular on $\scri^-$ and $\scri^+$, respectively.  Neglecting
back-scattering, a mode of the form (\ref{phi}) near $\scri^-$
takes, near $\scri^+$, the form
\begin{eqnarray}%
\varphi_{\Omega}(r,t) \approx {1 \over
(2\pi)^{3/2}(2\Omega)^{1/2} \, r} \; \e^{-i\Omega p(u)}\;.%
\label{phi+}%
\end{eqnarray}%
This can be regarded, approximately, as a mode of the type presented
in equation~(\ref{psi}), but now with $u$-dependent frequency
$\omega(u,\Omega)=\dot{p}(u)\;\Omega$, where a dot denotes
differentiation with respect to $u$. (Of course, this formula just
expresses the redshift undergone by a signal in travelling from
$\scri^-$ to $\scri^+$.)

In general we can expect a mode to be excited if the standard
adiabatic condition
\begin{equation}%
|\dot{\omega}(u,\Omega)|/\omega^2\ll 1%
\end{equation}%
does \emph{not\/} hold.  It is not difficult to see that this
happens for frequencies smaller than
\begin{equation}%
\Omega_0(u) \sim |\ddot{p}(u)|/\dot{p}(u)^2\;.%
\end{equation}%
One can then think of $\Omega_0(u)$ as a frequency marking, at each
instant of retarded time $u$, the separation between the modes that
have been excited ($\Omega\ll\Omega_0$) and those that are still
unexcited ($\Omega\gg\Omega_0$).

Moreover, Planck-scale modes (as defined on $\scri^-$) are excited
in a finite amount of time, even {\em before\/} the actual formation
of any trapped region. Indeed, they start to be excited when the
surface of the star is above the classical location of the horizon by
a proper distance $D$ of about one Planck length, as measured by
Schwarzschild static observers.  We can see this by observing that
the red-shift factor satisfies
\begin{equation}%
\left(1-2M/r\right)^{1/2} \sim
\omega/\Omega =\dot p(u) \sim
\kappa/\Omega_0\;,%
\end{equation}%
where $\kappa=(4M)^{-1}$ is the surface gravity.  This then implies
$(r-2M) \sim \kappa/\Omega_0^2$, where we have used $\kappa \, M\sim
1$. Hence
\begin{equation}%
D \sim (r-2M)\left(1-2M/r\right)^{-1/2}\sim 1/\Omega_0\;.%
\end{equation}%
Hence, the trans--Planckian problem has its roots at the very onset
of the formation of the trapping horizon. Furthermore, any complete
description of the semiclassical collapse cannot be achieved without
at least \emph{some\/} assumptions about trans--Planckian physics.

Of course, one can simply assume that there is a natural
Planck-scale frequency cutoff for effective field theory in curved
spacetimes. Although one cannot completely exclude this possibility,
we find that this way of avoiding the trans--Planckian problem is
perhaps worse than the problem itself, as it would automatically
also imply a shut-down of the Hawking flux in a finite (very small)
amount of time. This would eliminate the thermodynamical behaviour
of black holes, thus undermining the current explanation for the
striking similarity between the laws of black hole mechanics and
those of thermodynamics --- that they are, in fact, just the same
laws~\cite{davies}.

Moreover, such a ``hard cutoff" obviously corresponds to a breakdown
of Lorentz invariance at the Planck scale. If one is ready to accept
such a departure from standard physics, then it seems more plausible
(less objectionable?) to conjecture a milder breaking of Lorentz
invariance in the form of a modified dispersion relation, a
possibility explored in several works on the trans--Planckian
problem~\cite{Jacobson:1999zk}. While it is seemingly well
understood that the Hawking radiation would survive in this
case~\cite{US}, it is however less clear what effect such modified
dispersion relations might have on the possibility of forming a
(presumably frequency-dependent) trapping horizon, and indeed, on
the very definition of such a concept~\cite{Barcelo:2006yi}.

In what follows we shall adopt a conservative approach and stick, as
is usually done, to the standard framework of quantum field theory
in curved spacetime,  assuming its validity up to arbitrarily high
frequencies. Even in the presence of Lorentz violating effects, this
would remain a valid framework if, for example, the scale at which
Lorentz violations might appear was much higher than the Planck
scale~\cite{Jacobson}.

\subsection{Vacuum polarization}

The other difficulties of the standard scenario previously listed
have been linked by different authors to the presence of horizons
and of trapping regions in general. As we have previously discussed,
several departures from semiclassical gravity have often been called
for in order to solve these problems.  However, the specific
question we now want to raise here is rather different: Is the
scenario just described guaranteed to be the one actually realized
in semiclassical gravity? Or is it possible that semiclassical
gravity allows for alternative endpoints of gravitational collapse,
in which these problems are not present? In order to answer these
questions we look for possible semiclassical effects which could
modify the collapse before the very formation of a trapped region.

In any calculation of semiclassical collapse the choice of the
propreties of the matter involved (which will be encoded in the
characteristics of the classical SET) is, obviously, of crucial
importance. Normally the initial conditions at early times are
chosen so that one has a static star with any quantum field in their
``natural" vacuum state. As we have discussed, this will be
virtually indistinguishable from the Boulware vacuum state. In this
initial configuration we are sure that the RSET is practically zero
throughout spacetime, at least before the collapse is initiated. We
now want to inquire into the possibility that such a RSET becomes
non-negligible during the collapse.

In the standard semiclassical scenario, it is crucial that the
initial Boulware-like structure of the field modes at $\scri^-$ is
somehow ``excited'' by the collapse and converted into a Unruh-like
structure at both $\mathcal{H}^+$ and $\scri^+$ --- this is
necessary for compatibility with the presence of a trapping horizon.
In fact, if this excitation and conversion were not to be
sufficiently effective so as to to get rid of Boulware-like modes in
the proximity of the would-be horizon, then a potential obstruction
to the very formation of the horizon may arise. We know in fact that
in static geometries there is an intrinsic incompatibility between
the Boulware vacuum and the existence of a trapping horizon, as the
RSET near the horizon (in a simplified calculation in 1+1
dimensions) is found to be~\cite{dfu}
\begin{equation}%
\langle 0_B| \widehat T_{\hat \mu \hat
\nu}(r) |0_B\rangle_{\rm ren}\propto
-{1 \over
M^2} \; {1 \over 1-2M/r}
\left[
\begin{array}{cc}1&0 \\  0&1  \end{array}
\right]\,,%
\label{<T>}
\end{equation}%
where we work in an orthonormal basis. A similar result remains
valid in the more complicated $(3+1)$-dimensional case~\cite{scd}.
The important point is that the denominator vanishes at the horizon.
Hence the RSET acquires a divergent (and energy condition
violating~\cite{visser}) contribution. Note that the divergence {\em
is\/} present even if the components of the RSET are evaluated in a
freely-falling basis~\cite{scd}.  (To see that something {\em
intrinsic\/} is going on at the horizon it is sufficient to
calculate the scalar invariant $T_{\mu\nu}\;T^{\mu\nu} =
T_{\hat\mu\hat\nu} \; T^{\hat\mu\hat\nu}$, and to note that this
scalar diverges at the horizon.)

Of course the above result applies to a static spacetime, while we
are interested in investigating an intrinsically dynamical scenario,
which we moreover know, due to the Fulling--Sweeny--Wald
theorem~\cite{FSW}, should act in such a way as to avoid the above
divergence. We are hence interested in seeing the precise way in
which this happens, and in exploring whether it might leave a route
to possibly obtaining large, albeit finite, contributions to the
RSET at the onset of horizon formation.

\section{The RSET}%

In calculating the RSET in a dynamical collapse several choices must
be made. The major assumption is that we shall for the time being
restrict attention to $1+1$ dimensions, since then there is a
realistic hope of carrying out a complete analytic calculation.
Physically, this is not as bad a truncation as it at first seems,
since we can always view it as an $s$-wave approximation to full
$(3+1)$-dimensional problem, with at most a few actors of $r^{-2}$
being inserted at strategic places. (For instance, this analytic
approximation underlies the subsequent numerical calculation of
Parentani and Piran~\cite{pp}.)  A second significant choice we will
make is to specifically work in a regular coordinate system, in
particular, in Painlev\'e--Gullstrand
coordinates~\cite{pg,analogue}.  In regular coordinate systems
(where the matrix of metric coefficients is both finite and
non-singular), the values of the stress-energy-momentum  components
are direct and useful diagnostics of the ``size'' of the
stress-energy-momentum tensor.

\subsection{Preliminaries}%

With reference to the diamond-shaped conformal diagram of
Fig.~\ref{F:diamond}, we shall start by considering a set of affine
coordinates $U$ and $W$, defined on $\scri^-_\mathrm{left}$ and
$\scri^-_\mathrm{right}$ respectively. These coordinates are
globally defined over the spacetime and the metric can be written as
\begin{equation}
\g =  -C(U,W)\, \d U \, \d W\;.
\end{equation}
Given that we shall be concerned with events which lie outside of
the collapsing star on the right-hand side of our diagram, we can
also choose a second double-null coordinate patch $(u,W)$, where $u$
is taken to be affine on $\scri^+_\mathrm{right}$, in terms of which
the metric is
\begin{equation}
\g =  -\bar C(u,W) \, \d u \, \d W\;.
\end{equation}
Of course,
\begin{equation}
C(U,W) = {\bar C(u,W)/ \dot p(u) }\;,
\end{equation}
where $U=p(u)$ describes the coordinate transformation.
Then
\begin{equation}
\partial_U = \dot p^{-1} \, \partial_u\;.
\end{equation}
Furthermore, as long as we are outside  the collapsing star it is
safe to assume that a Birkhoff-like result holds, and take $\bar
C(u,W)$ as being that of a static spacetime.

Now for \emph{any\/} massless quantum field, the RSET (corresponding
to a quantum state that is initially Boulware) has
components~\cite{dfu, birrell-davies}
\begin{equation}
T_{UU} \propto C^{1/2}\, \partial_U^2\, C^{-1/2}\;,
\end{equation}
\begin{equation}
T_{WW} \propto C^{1/2}\, \partial_W^2\, C^{-1/2}\;,
\end{equation}
\begin{equation}
T_{UW} \propto R\;.
\end{equation}
The coefficients arising here are not particularly important, and
will in any case depend on the specific type of quantum field under
consideration.

The components $T_{WW}$ and $T_{UW}$ will necessarily be well
behaved throughout the region of interest; in particular they are
the same as in a static spacetime and are known to be regular. On
the contrary $T_{UU}$ shows a more complex structure due to the
non-trivial relation between $U$ and $u$. A brief computation yields
\begin{equation}
C^{1/2}\, \partial_U^2\, C^{-1/2} = {1\over\dot p^2} \left[
\bar C^{1/2}\, \partial_u^2\, \bar C^{-1/2} - \dot p^{1/2}\, \partial_u^2\, \dot p^{-1/2}
\right].
\label{T}
\end{equation}
The key point here is that we have two terms, one ($\bar C^{1/2}\,
\partial_u^2\, \bar C^{-1/2}$) arising purely from the static
spacetime outside the collapsing star, and the other ($\dot
p^{1/2}\, \partial_u^2\, \dot p^{-1/2} $) arising purely from the
dynamics of the collapse. If, and only if, the horizon is assumed to
form at finite time will the leading contributions of these two
terms cancel against each other --- this is the standard scenario.

Indeed the first term is exactly what one would compute from using
standard Boulware vacuum for a static star. As the surface of the
star recedes, more and more of the  static spacetime is
``uncovered'', and one begins to see regions of the spacetime where
the Boulware contribution to the RSET is more and more negative, in
fact diverging as the surface of the star crosses the horizon.

\subsection{Regular coordinates}%

To probe the details of the collapse, it is useful to introduce yet
a third coordinate chart --- a Painlev\'e--Gullstrand coordinate
chart $(x,t)$ in terms of which the metric
is~\cite{quasi-particle-prl, pg, analogue}
\begin{equation}
\g =  -c^2(x,t)\;  \d t^2 + [\,\d x - v(x,t)\,\d t\,]^2\;.
\end{equation}
This coordinate chart is particularly useful because it is regular
at the horizon, so that the finiteness of the stress-energy-momentum
components in this chart has a direct physical meaning in terms of
regularity of the stress-energy-momentum
\emph{tensor\/}.\footnote{These coordinates are also useful as they
allow to straightforwardly apply our calculations to acoustic
analogue spacetimes (provided one is in a regime in which one could
neglect the existence of modified dispersion
relations)~\cite{quasi-particle-prl,  analogue}. }
By setting the spacetime interval to zero, it is easy to see that
the null rays are given by
\begin{equation}
\d x = (\pm c + v) \; \d t.
\end{equation}

Although inside the collapsing star the metric can depend on $x$ and
$t$ in a complicated way, the geometry outside the surface of the
star is taken to be  static, so the functions $c$ and $v$ do not
depend on $t$.  Under these conditions we can integrate along the
history of an outgoing ray from an event $(t,x)$ just outside the
collapsing star to another event $(t_f,x_f)$ at asymptotic future
infinity $\scri^+_\mathrm{right}$:
\begin{equation}
t_f-t =  \int^{x_f}_x {\d x'\over c(x') + v(x')}\;.
\end{equation}
Assuming asymptotic flatness, $c(+\infty)=1$ and $v(+\infty)=0$, we
find for the $u$ null coordinate in the ``out'' region,
\begin{equation}
u := \lim_{t_f \to +\infty} \left(t_f-x_f\right)
= t - \int^x {\d x' \over c(x')+v(x')}\;.
\label{u}%
\end{equation}
Hence, denoting partial derivatives by subscripts:
\begin{equation}
U_x = \dot p(u) \, u_x= - {\dot p(u)\over c(x)+v(x)};
\end{equation}
\begin{equation}
\qquad U_t = \dot{p}(u)\,u_t = \dot p(u)\;.
\end{equation}
In contrast, along an incoming ray leaving asymptotic past infinity
$\scri^-_\mathrm{right}$ at an event $(t_i,x_i)$ and remaining
outside the star,
\begin{equation}
t-t_i = - \int_{x_i}^x {\d x'\over c(x') - v(x')}\;,
\end{equation}
so we have, for the $W$ null  coordinate:
\begin{equation}
W := \lim_{t_i\to -\infty}\left(t_i+x_i\right)
= t + \int^x \frac{\d x'}{c(x')-v(x')}\;.
\label{W}%
\end{equation}
Hence
\begin{equation}
W_x = {1\over c(x) - v(x)}; \qquad W_t = 1.
\end{equation}

In addition, by substituting and comparing coefficients of the line
element, it is easy to see that the $(U,W)$ and  $(x,t)$ coordinates
are related by
\begin{equation}
U_t = - (c+v)\, U_x\;,
\end{equation}
\begin{equation}
W_t =  (c-v)\, W_x\;,
\end{equation}
and
\begin{equation}
C(x,t) = -{1\over U_x\, W_x}\;.
\end{equation}

\begin{widetext}
Therefore the components of the RSET can be calculated in any of the
equivalent forms:
\begin{eqnarray}
T_{tt} &=& U_t^2 \, T_{UU} + 2\, U_t \, W_t \, T_{UW} + W_t^2 \, T_{WW} \\
&=& (c+v)^2\, U_x^2\, T_{UU} - 2\, (c^2-v^2)\, U_x\, W_x\, T_{UW} + (c-v)^2\, W_x^2\, T_{WW} \\
&=& \dot p^2 \, T_{UU} - 2\, \dot p \, T_{UW} + T_{WW}\;;\\
&&\nonumber\\
T_{tx} &=& U_t \, U_x\, T_{UU} + \left(U_t \, W_x + U_x \, W_t\right) \, T_{UW} + W_t \;W_x\; T_{WW} \\
&=& -(c+v)\, U_x^2\, T_{UU} - 2\, v\, U_x\, W_x\, T_{UW} + (c-v)\, W_x^2\, T_{WW} \\
&=&  -{\dot p^2\over c+v} \,T_{UU} + {2\,\dot p\, v\over c^2- v^2} \, T_{UW} + {1\over c-v} \, T_{WW}\;;\label{tx}\\
&&\nonumber\\
T_{xx} &=& U_x^2\, T_{UU} + 2\, U_x\, W_x\, T_{UW} + W_x^2\, T_{WW} \\
&=&
{\dot p^2\,\over (c+v)^2}\,T_{UU} - 2\, {\dot p\over c^2-v^2}\, T_{UW} + {1\over (c-v)^2}\, T_{WW}\;. \label{xx}
\end{eqnarray}
Some of these formulae are more useful for calculating the static
Boulware contribution, others are more useful for calculating the
dynamical contribution. Since $c+v\to0$ at a horizon, while
$c-v\to2c$ is regular, this is enough to guarantee that the $T_{tt}$
and $T_{tx}$ components of the RSET are always better behaved (less
divergent) than the $T_{xx}$ component. Note that no divergence can
arise from the terms proportional to $T_{WW}$. \vskip 0.2 cm
\end{widetext}

Equations (\ref{u}) and (\ref{W}) also allow us to express the
derivative with respect to $u$ in terms of those with respect to the
regular coordinates $x$ and $t$:
\begin{equation}%
\partial_u=\frac{c+v}{2\,c}\,\partial_t-\frac{c^2-v^2}{2\,c}\,\partial_x\;.%
\label{du}%
\end{equation}%

\subsection{Calculation assuming normal horizon formation}%

Hereafter, we shall for simplicity restrict our attention to the
case $c(x)\equiv 1$.  Placing the horizon at $x=0$ for convenience,
we can write the asymptotic expansion
\begin{equation}%
v(x) \approx -1 + \kappa\, x + \kappa_2\, x^2 + \cdots\;,%
\label{v}%
\end{equation}%
where $\kappa$ can be identified with the surface
gravity~\cite{quasi-particle-prl,analogue}.

Consider first the static Boulware term in equation (\ref{T}). We
have (placing the horizon at $x=0$ for convenience)
\begin{equation}
\bar C = -\frac{\dot{p}}{U_x\,W_x} = -{1\over u_x\, W_x} = 1 -
v(x)^2 \approx 2\, \kappa\, x\;.
\end{equation}
The relevant derivative in $\partial_u$ is then that with respect to
$x$, and we can write
\begin{eqnarray}
\bar C^{1/2}\, \partial_u^2\, \bar C^{-1/2} &\approx&
 (2\,\kappa\, x)^{1/2}\,  \kappa\, x\, \partial_x \left( \kappa\, x\, \partial_x (2\,\kappa\, x)^{-1/2} \right)\nonumber \\
 &=& \kappa^2/4\;.
\end{eqnarray}
In fact, keeping the subleading terms one finds
\begin{equation}
\bar C^{1/2}\, \partial_u^2\, \bar C^{-1/2} = {\kappa^2\over4} + \mathcal{O}(x^2).
\end{equation}
By equations (\ref{tx}) and (\ref{xx}), it is clear that because of
the constant term $\kappa^2/4$, the components $T_{tx}$ and $T_{xx}$
of the RSET contain contributions that diverge as $x^{-1}$ and
$x^{-2}$, respectively, as $x\to 0$. (The sub-leading terms lead to
finite contributions of order $\mathcal{O}(x)$ and $\mathcal{O}(1)$
respectively.)

In counterpoint, assuming horizon formation, let us now calculate
the dynamical contribution to the RSET ($\dot p^{1/2}\,
\partial_u^2\, \dot p^{-1/2} $). It is well known that any
configuration that produces a horizon at a finite time $t_{\rm H}$
leads to an asymptotic (large $u$) form
\begin{eqnarray}%
p(u)\approx U_{\rm H} - A_1\; \e^{-\kappa u}\;,
\label{standard-p}
\end{eqnarray}%
where $U_H$ and $A_1$ are suitable constants. Taking into account
the asymptotic expression (\ref{v}) for $v(x)$ near $x=0$, it is
very easy to see that the potential divergence at the horizon due to
the static term is exactly cancelled by the dynamical term. In this
way we have recovered the standard result that the RSET at the
horizon of a collapsing star is regular.

However, the previous relation is an asymptotic one, and for what we
are most interested in (the value of the RSET close to horizon
formation) it is important to take into account extra terms that
will be subdominant at late times. Indeed, we can describe the
location of the surface of a collapsing star that crosses the
horizon at time $t_{\rm H}$ by
\begin{equation}
x= r(t) -2M = \xi(t) = -\lambda (t-t_{\rm H})+\cdots\;,
\end{equation}
where the expansion makes sense for small values of $|t-t_{\rm H}|$,
and $\lambda$ represents the velocity with which the surface crosses
the gravitational radius.  Let $t_0$ be the time at which a
right-moving light ray corresponding to  null coordinates $u$ and
$U$ crosses the surface of the star.  Then on the one hand
\begin{eqnarray}%
t_f-t_0=\int_{\xi(t_0)}^{x_f} \frac{\d x'}{1+v(x')}\;,
\end{eqnarray}%
which for $t_0\approx t_{\rm H}$ (implying $r(t_0)\approx 2M$) can
be approximated by
\begin{eqnarray}%
u \approx \left(t_0-t_{\rm H}\right) - \frac{1}{\kappa}\ln \left(-\lambda \left(t_0-t_{\rm H}\right)\right) +C_1\;,
\end{eqnarray}%
so that
\begin{equation}
t_0 - t_{\rm H} \approx
C_2 \,{\e^{-\kappa u}\over\lambda} + \cdots
\label{t0-tH}
\end{equation}
On the other hand, since $U(t_0)$ is simply some regular function, we have
\begin{equation}
U(t_0) = U_{\rm H} + U'_{\rm H} \; (t_0-t_{\rm H}) + {U''_{\rm H}\over2} \; (t_0-t_{\rm H})^2 +\cdots
\label{Ut0}
\end{equation}
Inserting (\ref{t0-tH}) into (\ref{Ut0}) we obtain an asymptotic expansion
\begin{eqnarray}%
p(u) = U_{\rm H} - A_1 \; \e^{-\kappa u} + {A_2\over 2}\;  \e^{-2\kappa u}
+ {A_3\over 3!} \; \e^{-3\kappa u}+ \cdots\;\;
\label{expp}
\end{eqnarray}%
which it is useful to write as
\begin{equation}
p(u) = U_{\rm H} - F(\e^{-\kappa u})\;,
\label{pF}
\end{equation}
where $F$ is a regular function such that $F(0)=0$.  Then
\begin{eqnarray}
\dot p^{1/2}\, \partial_u^2\, \dot p^{-1/2}
&=&
-{1\over2} {\dddot p\over\dot p} + {3\over4} \left({\ddot p\over\dot p}\right)^2
\nonumber\\
&=&
\frac{\kappa^2}{4} + \left[ -{1\over2} {F'''\over F'} + {3\over4} \left({F''\over F'}\right)^2 \right]\kappa^2\, \e^{-2\kappa u}
\nonumber\\
&=&
\frac{\kappa^2}{4} + \left[ -{1\over2} {A_3\over A_1} + {3\over4} \left({A_2\over A_1}\right)^2 \right]\kappa^2\, \e^{-2\kappa u}
\nonumber\\
&&
\qquad + \vphantom{{1\over4}} \mathcal{O}\left( \e^{-3\kappa u} \right)\;.
\label{dyn}
\end{eqnarray}
The point is that this has a universal contribution coming from the
surface gravity, plus messy subdominant terms that depend on the
details of the collapse.  It is important to note, however, that the
corresponding additional contributions to the RSET are finite, in
contrast to the one associated with the first term. Indeed, for
small values of $x$,
\begin{equation}
u \approx t-\frac{1}{\kappa}\,\ln x +\mbox{const}\;,
\end{equation}
so
\begin{equation}
\e^{-\kappa u}\propto x\;\e^{-\kappa t}\;,
\end{equation}
and so the second term in the right-hand side of equation
(\ref{dyn}) is $\mathcal{O}(x^2)$, and by equation (\ref{xx}) gives
an $\mathcal{O}(1)$ contribution to $T_{xx}$ that does not depend on
$x$, but depends on time as $\e^{-2\kappa t}$. In addition, from a
comparison of equations (\ref{t0-tH})--(\ref{expp}) we see that
\begin{equation}
{A_2\over A_1} \propto {1\over\lambda}\;,\qquad {A_3\over A_1}\propto {1\over \lambda^2}\;,
\end{equation}
so the leading subdominant term in the RSET is inversely
proportional to the square of the speed with which the surface of
the star crosses its gravitational radius. In particular, at horizon
crossing, that is at $t=t_{\rm H}$, the value of the RSET can be as
large as one wants provided one makes $\lambda$ very small. This
would correspond to a very slow collapse in the proximity of the
trapping horizon formation.  Thus, there is a concrete possibility
that (energy condition violating) quantum contributions to the
stress-energy-momentum tensor could lead to significant deviations
from classical collapse when a trapping horizon is just about to
form.

\subsection{Calculation assuming asymptotic horizon formation}%

Another interesting case one may want to consider is one in which
the horizon is never formed at finite time, but just approached
asymptotically as time runs to infinity. In particular, in
reference~\cite{quasi-particle-prl} it was shown that collapses
characterized by an exponential approach to the horizon,
\begin{equation}
r(t) = 2M + B \e^{-\kappa_{\rm D} t}\;,
\end{equation}
lead to a function $p(u)$ of the form
\begin{eqnarray}%
p(u)=U_{\rm H} - A_1 \e^{-\kappa_{\rm eff} u}\;,
\end{eqnarray}%
where $\kappa_{\rm eff}$ is half the harmonic mean between $\kappa$
and the rapidity of the exponential approach $\kappa_{\rm D}$,
\begin{equation}
\kappa_\mathrm{eff} = {\kappa \; \kappa_{\rm D}\over \kappa + \kappa_{\rm D}}\;,
\end{equation}
so that one always has $\kappa_{\rm eff}<\kappa$. In this case, the
calculation of the dynamical part of the RSET leads to exactly the
same result that when using expression (\ref{standard-p}), modulo
the substitution of  $\kappa$ by $\kappa_{\rm eff}$. However, the
non-dynamical part of the RSET remains unchanged. This implies that
now, at leading order
\begin{eqnarray}%
\hbox{RSET}(x \approx 0)
&\approx& {1 \over \kappa^2 x^2}\left(\kappa_{\rm eff}^2-\kappa^2\right)\nonumber\\
&=&%
-\frac{\kappa\left(2\,\kappa_{\rm D}+\kappa\right)}{\left(\kappa_{\rm D}+\kappa\right)^2\,x^2}\;,
\end{eqnarray}%
which obviously diverges in the limit $x \to 0$.  We stress that
this result does not contradict the Fulling--Sweeny--Wald
theorem~\cite{FSW}, as the calculation applies only outside the
surface of the star (\ie, for $x\geq\xi(t)$), and so the divergence
appears only at the boundary of spacetime.  Nevertheless,
particularizing to $x=\xi(t)$, this again indicates that there is a
concrete possibility that energy condition violating quantum
contributions to the stress-energy-momentum tensor could lead to
significant deviations from classical collapse when a trapping
horizon is on the verge of being formed.

\subsection{Physical insight}%

The key bits of physical insight we have garnered from this
calculation are:
\begin{itemize}
\item
In the standard collapse scenario the regularity of the RSET at
horizon formation is due to a subtle cancelation between the
dynamical and the static contributions.
\item
Contributions that can be neglected at late times can indeed
be very large at the onset of horizon formation. The actual
value of these contributions depends on the rapidity with which
the configuration approaches its trapping horizon.
\item
Once the horizon forms, the above contributions will be exponentially
damped with time. However, the analysis of the configuration
that approches horizon formation asymptotically tells
us that, while horizon formation is delayed, there are
contributions that will keep growing with time.
\end{itemize}

Hence apparently the RSET can acquire large (and energy condition
violating~\cite{visser}) contributions when a collapsing object
approaches its Schwarzschild radius, depending on the details of the
dynamics. The final lesson to draw from this part of our
investigation is that not all the classical matter configurations
compatible with the formation of a trapping horizon in classical
general relativity necessarily lead to the same final state when
semiclassical effects are taken into account. In particular, for
classical collapses that exhibit a slow approach to horizon
formation, our calculation indicates that there will be a large
(albeit always finite in compliance with \cite{FSW}) contribution
from the RSET, a contribution which can potentially lead the
semiclassical collapse to classically unforeseeable end points.  For
these reasons we wish next to further explore the alternative
situation in which the horizon is only formed asymptotically.

\section{A quasi-black hole scenario}%

The history of the confrontation between general relativity and
quantum physics has already shown several times that the quantum
mechanical effects in matter can prevent the formation of black
holes in situations in which classically such formation would seem
unavoidable. Without quantum mechanics, objects such as white dwarfs
and neutrons stars would have never been predicted in the first
place.  Similarly, in this paper we have seen that if for any reason
the collapse of the matter forces it into some (metastable) state in
which horizon formation is approched sufficiently slowly, then large
quantum vacuum effects could prevent the very formation of a
trapping horizon. The resulting object could then be considered the
most compact and quantum mechanical kind of star. These objects,
which we shall tentatively call ``black stars'',\footnote{Newtonian
versions of ``black stars'',  more often called ``dark stars'', have
a very long history in astrophysics, dating back to
Michell~\cite{Michell} and Laplace~\cite{Laplace}. For recent
commentary on the historical connections between Michell, Cavendish,
and Laplace, see~\cite{Lynden-Bell}.} would be supported by a form
of quantum pressure of universal nature, being characterized only by
their closeness to the formation of a trapping horizon.

Lacking an understanding of the physics of matter at densities well
beyond that characterizing neutron stars, we cannot reliably assert
anything about the stability of black stars. However, the first
motivation for our investigation was to see whether semiclassical
physics can allow for compact objects closely mimicking black hole
features, including Hawking radiation, without incurring in the same
problems plaguing the standard scenario. In this sense, static
configurations do not seem viable candidates as the absence of a
trapping horizon together with the staticity prevents any
possibility of emission of a Hawking flux.\footnote{It is perhaps
worthwhile to stress here that such static black stars do not belong
to the class of objects known in the literature as gravastars, (at
least not without the addition of considerable extra assumptions),
given than the former are compact aglomerates of matter while the
latter have a de~Sitter-like interior~\cite{gravastar}.}  On the
other side, evolving configurations that continue to asymptotically
approach their would-be horizon\footnote{This approach could be
completely monotonic or have oscillating components. These
oscillations can also produce burst of radiation at the Hawking
temperature~\cite{thooft}.} would produce quantum radiation at late
times.

In order for such a scenario to be realized in nature one can
speculate that in some cases, once matter has slowed down the
collapse so allowing for the piling up of a sizeable RSET, the
latter would not be able to completely stop the collapse, but would
instead lead to an evolving configuration where every layer of the
collapsing star would lie very close to where the classical horizon
of the matter inside it would be located, continually and
asymptotically approaching it. We can call this object a
``quasi-black hole''.

In order to know exactly how the star asymptotically approaches the
horizon in this scenario, one should solve Einstein's semiclassical
field equations with back-reaction --- obviously a very difficult
task. Without the result of such an explicit calculation, it is
nevertheless reasonable to conjecture that the approach can either
follow a power law, or be exponential with a timescale
$1/\kappa_{\rm D}$, say. The case of a power law seems, however,
uninteresting for our purposes, because it would not lead to a
Planckian emission~\cite{quasi-particle-prl}.  On the contrary, an
exponential approach is associated with the emission of radiation at
a modified temperature $T=\left(2\pi/\kappa+2\pi/\kappa_{\rm
D}\right)^{-1}$~\cite{quasi-particle-prl}.  At least for
astrophysical black holes, it is also reasonable to think that
$\kappa_{\rm D}\gg \kappa$ at the beginning of the evaporation
process, so that $T \approx \kappa/2\pi$, indistinguishable from the
standard Hawking temperature. During evaporation $\kappa$ increases
so, in the long run, $T$ is determined by $\kappa_{\rm D}$ and tends
to zero.  Hence we could in principle have a ``graceful exit'' from
the evaporation process; that is, one could avoid the standard
run-away endpoint. Meanwhile, the evaporation could be visualized as
a continuous chasing between the surface of the star and its
(receding) Schwarzschild radius.

Indeed, possibilities for such a never-ending collapse were already
envisaged in 1976, soon after the discovery of Hawking
radiation~\cite{boulware2, gerlach} and have been recently proposed
again~\cite{Vachaspati:2006ki} (although via different back-reaction
mechanisms). It is important, however, to understand that in the
quasi-black hole scenario we discuss here the Hawking flux only
affects late-time evolution, and is {\em not\/} the agent that
prevents horizon formation in the first place.  The initial
slow-down of the collapse is in this case due to matter-related high
energy physics. This provides the time necessary for the vacuum
polarization  to grow and finally modify  the evolution of the
collapse toward an asymptotic regime.

Of course, the state at $T=0$ is reached only after a very long time
(for typical estimates of the evaporation timescale, see
reference~\cite{birrell-davies}), so according to this scenario a
collapsing star forms an object that, for a long period, is
indistinguishable from a standard black hole, further justifying our
nomenclature of ``quasi-black hole''.  This object would still
evaporate with a Planckian spectrum~\cite{quasi-particle-prl}, but
(since there is no event horizon) it would not be truly ``thermal''
(the quantum state is indeed a squeezed state~\cite{squeezed}),
hence there would be no information-loss problem.  The partners of
the particles emitted towards infinity, instead of being accumulated
inside the trapping horizon as in the standard scenario, would now
simply be emitted with a (significant) temporal delay. The radiation
received at one instant of time would be correlated with that
arriving some time later, so all the information would be recovered
in the resulting radiation.

How does back-reaction work in this scenario?  During the late time
asymptotic collapse, two processes unfold at the same time: (1) the
energy associated with vacuum polarization becomes more and more
negative; (2) radiation is emitted towards infinity.  During a time
interval $\Delta u$ as measured on $\scri^+$, an arrival of energy
$\Delta E_{\rm rad}>0$ is recorded by observers at infinity.
Correspondingly, vacuum polarization leads to an extra energy
$\Delta E_{\rm vac}<0$ (due to the fact that the star becomes more
compact), so the Bondi mass of the object decreases by an amount
$|\Delta E_{\rm vac}|$.  By energy conservation, one expects that
$\Delta E_{\rm vac}=-\Delta E_{\rm rad}$, so the emission of
radiation is balanced by the increase of vacuum polarization nearby
the central object.  This balance makes the Bondi mass of the object
decrease {\em as if\/} it were taken away by radiation, eventually
reducing to zero as $T\to 0$. Note that the expression (\ref{<T>})
for the RSET can be rewritten in such a way as to exhibit the fact
that vacuum polarization corresponds to the \emph{absence\/} of
black-body radiation at the temperature $T=(8\pi
M)^{-1}$~\cite{scd}. Although this does not constitute a proof, it
is a strong plausibility argument in favour of the energy balance
between radiation and vacuum polarization.  Also, it strongly
suggests that the asymptotic approach to the would-be horizon must
be of the exponential type, rather than a power law. Indeed, since a
power law would not lead to a Planckian emission, it would be hard
to reconcile it with the result presented in reference~\cite{scd}.

Thus, provided that trapping horizons do not form, we have
described a plausible scenario for the progressive collapse and
evaporation of quasi-black holes. However, the end point of this
process seems to still share a problem with the standard scenario:
The apparent accumulation of baryon number within the collapsing
object~\cite{boulware2}. The least massive baryon one can find is
the proton. Baryon number is conserved in all experiments realized
up to now, and in particular, the proton has been found to be
stable (nevertheless, Grand Unification Theories predict it should
eventually disintegrate into leptons). In the standard paradigm for the
evaporation of a black hole, the trapping horizon and its
surroundings is an empty region of spacetime. Therefore, there is
only one physical quantity characterizing the quantum emission:
The value of its Hawking temperature.  For a standard evaporating
black hole to be able to nucleate a proton-antiproton pair, it
seems necessary that it reaches a temperature larger than $\sim
10^{13}$ K, or equivalently, a tiny mass of less than $\sim
10^{38}m_p$, where $m_p$ is the mass of one proton. However, for
example, a black hole having initially one solar-mass would
contain a baryon number of around $\sim 10^{57}$. During the
evaporation it would conserve this baryon number till it reaches a
Bondi mass of $\sim 10^{38}m_p$. But then, even emitting all its
remaining energy in the form of baryons (with emission in the form
of protons being the most efficient way of removing baryon
number), it would end up either: (1) leaving an almost massless
relic having a baryon number of $\sim 10^{57}-10^{38}\sim
10^{57}$ (a rather peculiar state); or (2) completely evaporating
producing an enormous violation of baryon-number conservation.

The quasi-black hole scenario, however, adds one extra ingredient to
the previous discussion:  The would-be horizon and its surroundings
is now not an empty region of spacetime.  In the vicinity of the
would-be horizon there is always matter progressively being
compressed.  This fact could significantly affect the way the
quasi-black hole radiates its energy.  For example, an upper bound
for the average density of a solar mass quasi-black hole is given by
that of the corresponding black hole $\sim1/M^2\sim 10^{19} \, {\rm
kg}/{\rm m}^3$ (a few times bigger than that of a typical neutron
star).  At these densities and higher, it is quite plausible that
new particle physics effects could come into play and deplete the
baryon number much more efficiently than the evaporation
process.\footnote{Of course, for very massive quasi-black holes such
effects will be negligible for a very long time, but will eventually
become important as the Bondi mass is decreased by the combined
effect of Hawking radiation and vacuum polarization.}

Up to this point we have only considered spherically symmetric
configurations. However, current observations tell us that most of
the observed black hole candidates have a high rate of rotation,
sometimes very close to extremality~\cite{rotation}. Hence, for a
quasi-black hole scenario to be a feasible description of these
objects, it would be necessary to generalize our proposal to
rotating configurations. Given the complexity of the vacuum
structure around rotating black holes~\cite{ottewill} it is very
difficult to have a precise proposal in this sense. However, we know
that any rotating object possessing an ergoregion but not a horizon
would be highly unstable~\cite{cardoso}. Hence we expect that any
viable model of a rotating quasi-black hole should be characterized
by a matter distribution extending up to the outer boundary of the
ergoregion.

The fact that most of the progenitors of the observed black hole
candidates are characterized by supercritical rotations ($J>M^2$,
where $J$ is the angular momentum of the progenitor) is often used
as evidence of the validity of the cosmic censorship conjecture.  It
is interesting to note that if such conjecture holds for standard
general relativity it would also be effective in preventing
super-critical quasi-black holes. In order to understand this point
it is enough to realize that a generalization of the calculation of
this article to more general metrics allowing for extremality (\eg,
Reissner--Nordstr\"om, Kerr, ...) would still imply a pile up of the
RSET in proximity of the ``would-be horizon'' {\em if and only if\/}
such a horizon can form in the first place. That is, a large
quantum-induced RSET can arise only if the collapsing object has
already shed the extra charges (\eg, electric charge or angular
momentum) so as to be subcritical in proximity of the horizon
crossing. So supercritical configurations are likely to be
unaffected by the vacuum polarization and behave as in classical
general relativity. On the contrary sub-critical configurations will
develop (or not develop) trapping horizons according to the details
of the dynamics.

\section{Conclusions}%

Quantum physics imposed upon the description of the collapse of
astrophysical objects in situations that would classically lead to
black hole formation could unexpectedly lead to observable effects
at early times, when the trapping horizon is about to form. In
particular we have shown that before forming a trapping horizon,
trans--Planckian modes are excited.  Hence, whether the trapping
horizon forms or not depends critically on assumptions concerning
the net effect of any trans--Planckian physics that might be at
work.

Assuming that quantum field theory holds unmodified up to
arbitrarily high energy (as is commonly done in most of the extant
literature) we have shown that there can be large deviations from
classical collapse scenarios, if the latter do allow in the first
place a piling up of vacuum energy. Most of the classical collapse
scenarios so far considered do not allow for such a piling up, due
to their intrinsic rapidity. In this sense the prediction of horizon
formation in many of these models~\cite{pp, massar} seems completely
correct.

We have argued however, that alternative classical collapse
scenarios in which horizon formation is approached in a slow manner
are not only foreseeable, but possibly natural in more realistic
situations. If this is indeed the case one then would have to add a
new class of compact, horizonless, objects (possibly the most
compact objects apart from black holes themselves) to the
astrophysical bestiary: the black stars.

In the final part of this work we have then considered a particular
subclass of these objects, the quasi-black holes, which could
closely mimic all the most relevant features of black hole physics,
while avoiding at the same time most of its intrinsic problems (such
as singularities, the information paradox, and the question of the
end point of Hawking evaporation).

Summarizing, the quasi-black hole scenario for collapse and
evaporation is the following one (see Fig.\ \ref{F:qbh}):
%
\begin{figure}[htb]%
\vbox{ \hfil \scalebox{0.60}[0.60]{{\includegraphics{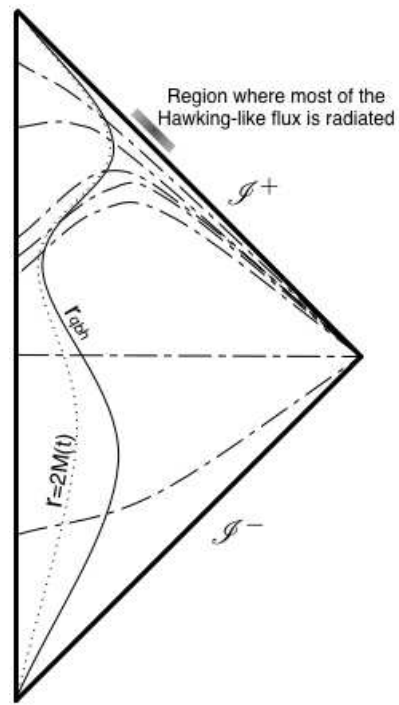}}}\hfil}%
\bigskip%
\caption{Conformal diagram of the spacetime for a quasi-black hole.
The solid line represents the surface of the collapsing object; the
dotted line is at $r=2M(t)$, where $M(t)$ is the instantaneous mass
of the object as measured from $\scri^+$; dashed lines correspond to
(Schwarzschild) $t=\mbox{const}$ hypersurfaces. The period of
evaporation appears short because of a distortion induced by the
representation, but actually corresponds to a very long lapse of
time, as one can see from the fact that the lines at
$t=\mbox{const}$ crowd around it. This diagram is compatible with
current astrophysical observations of gravitationally active
collapse products.}
\label{F:qbh}%
\end{figure}%
%
As a star of mass $M$ implodes we conjecture that its matter will
try to adjust in new, possibly unstable, configurations so to reach
a new equilibrium against gravity. If there is ever a significant
slowing down of the collapse, for any reason whatsoever, then this
allows the vacuum polarization to progressively grow, and further
slow down the approach to trapping horizon formation. Provided such
an approach is asymptotic with an exponential law controlled by a
timescale $1/\kappa_{\rm D}$, then the quantum radiation produced
during this process is still Planckian, with a temperature
$T=\left(2\pi/\kappa+2\pi/\kappa_{\rm D}\right)^{-1}$, where
$\kappa$ is inversely proportional to the total Bondi mass $M+E_{\rm
vac}$ of the star~\cite{quasi-particle-prl}.  For a long time,
$|E_{\rm vac}|\ll M$ and $\kappa\ll\kappa_{\rm D}$, so
$T\approx\kappa/2\pi$ and (from the point of view of an external
observer) the object is essentially indistinguishable from a
standard evaporating black hole. The emission of radiation is
accompanied by an increase in vacuum polarization, that
progressively diminishes the Bondi mass of the star, so the would-be
horizon shrinks and is never crossed by the matter configuration.
When the Bondi mass has become sufficiently small, $1/\kappa$ is
negligible and the temperature is approximately equal to
$\kappa_{\rm D}/2\pi$.  This quantity is also decreasing, because
back-reaction is in fact slowing down collapse, so the temperature,
after reaching a maximum value, decreases and approaches zero.

We do not yet have a definitive proposal as to the end-point of the
evaporation process. This could only be achieved by understanding
the physics of baryon nucleation in the presence of high-density
states of matter. The end state of the evaporation could correspond
to a zero-temperature relic\footnote{Note that the nature of such a
relic would be quite different from that of a standard black hole
remnant, because the relic could be regarded just as a peculiar case
of a very compact star. For this reason, the usual issues related to
remnants (like the compatibility with CPT invariance or their
capacity for storing information) are not present in this scenario.}
with vanishing Bondi mass (hence would at large distances be
gravitationally inert), with an inner structure formed by a core
with mass $\sim M$ and a non-vanishing baryon number, immersed into
a cloud of polarized vacuum with negative energy $E_\mathrm{vac}
\sim -M$. Alternatively, it might correspond to plain vacuum.

\section*{Acknowledgments}%

We are grateful to Daniele Amati, Larry Ford, Ted Jacobson, John
Miller, Jos\'e Navarro--Salas, Tom Roman, and Bernard Whiting for
critically reading a preliminary version of this paper and for
stimulating discussions.  We would like to also thank Renaud
Parentani and Robert Wald for their comments.  CB\ has been funded
by the Spanish MEC\ under project FIS2005-05736-C03-01 with a
partial FEDER\ contribution.  CB\ and SL\ are also supported by an
INFN-MEC\ collaboration.  MV\ was supported by a Marsden grant
administered by the Royal Society of New Zealand, and also wishes to
thank both SISSA/ISAS\ (Trieste) and IAA\ (Granada) for hospitality.



\end{document}